# Towards quantitative tissue absorption imaging by combining photoacoustics and acousto-optics

K. Daoudi, and W. Steenbergen[*]

*Biomedical Photonic Imaging group, MIRA Institute for Biomedical Technology and Technical Medicine, University of Twente, PO Box 217, 7500 AE Enschede, The Netherlands*

[*]*w.steenbergen@utwente.nl*

We propose a strategy for quantitative photoacoustic mapping of chromophore concentrations that can be performed purely experimentally. We exploit the possibility of acousto-optic modulation using focused ultrasound, and the principle that photons follow trajectories through a turbid medium in two directions with equal probability. A theory is presented that expresses the local absorption coefficient inside a medium in terms of noninvasively measured quantities and experimental parameters. Proof of the validity of the theory is given with Monte Carlo simulations.

## 1. Introduction

Photoacoustic imaging (PAI) is a biomedical imaging modality of rapidly increasing impact. PAI provides images of turbid media, based on ultrasonic waves created inside the medium by local absorption of short light pulses. The underlying effect is photoelastic energy conversion: local absorption of short light pulses leads to a mechanical stress which relaxes by the emission of ultrasonic waves. From these waves, the initial stress distribution can be reconstructed, which leads to three dimensional imaging of optical absorption. PAI has been elaborated into various implementations, mainly leading to devices for various imaging depths and imaging resolutions [1].

While PAI provides images about optical absorption it is inherently unable to render quantified images of the absorption coefficient. This is related to the fact, that under the conditions of heat and stress confinement, the initial stress distribution $\sigma(x, y, z)$ depends on the distribution of absorbed energy density $E_a$ as $\sigma = \Gamma E_a = \Gamma \mu_a F = \Gamma F c \varepsilon_a$ with $\Gamma$ the Grüneisen parameter, $\mu_a$ the local absorption coefficient, F the local fluence, $c$ the concentration of the chromophore (a single chromophore is assumed) and $\varepsilon_a$ its molar absorption. All above quantities vary with position in the tissue. The quantification problem in photoacoustics is to decompose the reconstructed initial stress distribution $\sigma_0(x, y, z)$ into fluence F and absorption coefficient $\mu_a$.

Efforts have been made to solve this problem by injecting an absorbing dye with known concentration [2], combining photoacoustics with the application of a model of light transport [3-8], optical determination of bulk optical properties [9], use of the property that often fluence and optical absorption vary at widely different length scales [10], and combining photoacoustics with diffuse optical tomography, using iterative computational models to achieve quantitative absorption and scattering images[11, 12].

In this letter we theoretically describe and validate methodology to determine the local absorption coefficient by combining PAI with acousto-optic modulation. We demonstrate that the local absorption coefficient can be established in a purely experimental manner, based on noninvasively measured quantities, and with the only parameters being the geometrical properties of the acousto-optic modulation and detection.

## 2. Theory

Consider points $i$ and $j$ inside or on the surface of a turbid medium, as depicted in Fig.1. On injection in point $i$ of light at power $P_i$, through an aperture placed in point $j$ with area $A_j$ and solid opening angle $\Omega_j$ an optical power is detected of

$$P_{ij} = A_j \Omega_j P_i \Pr(i, j) \qquad (1)$$

Here $\Pr(i, j)$ is the probability per unit aperture area and per unit solid angle that a photon starting in $i$ will cross an aperture at point $j$, following any possible photon trajectory. The fluence rate $\Phi_{ij}$ at point $j$ can be written

$$\Phi_{ij} = 4\pi P_i \Pr(i, j) \qquad (2)$$

The expressions for $P_{ij}$ and $\Phi_{ij}$ are in mutual agreement for an assumed isotropic radiance in point $j$. In an analogous manner, a pulse with pulse energy $E_{p,i}$ applied at surface point $i$ will generate fluence

$$F_{ij} = 4\pi E_{p,i} \Pr(i, j) \qquad (3)$$

Probability $\Pr(i, j)$ is affected by the unknown optical properties of that part of the medium that is interrogated by the light travelling from $i$ to $j$. A key concept of our method is that we exploit the principle that all photon trajectories contributing to $\Pr(i, j)$ can be followed in both directions with equal probability, hence $\Pr(i, j) = \Pr(j, i)$. Photon reversibility has been recently used in an extreme form by Xu et al. [13] for refocusing ultrasound-tagged photons in a scattering medium to the point of tagging, a process which involves both the phase and amplitude of the light. In our use of light reversibility we only use the intensity aspect of light and no phase conjugation is needed.

These principles are now applied to the medium depicted in Fig. 2, with surface points 1 and 3 and a volume around an internal point 2 with absorption coefficient $\mu_{a,2}$. Under the condition of stress confinement, the absorbed energy density at point 2 is

$$E_{a,i2} = F_{i2}\mu_{a,2} = 4\pi E_{p,i}\Pr(i,2)\mu_{a,2} \quad (4)$$

with $i=1,3$ for injection of light at points 1 and 3, respectively. Here we used Eq. (3) relating the internal fluence to the injected pulse energy. This leads to local stresses

$$\sigma_{2i} = \Gamma F_{i2}\mu_{a,2} = 4\pi\Gamma E_{p,i}\Pr(i,2)\mu_{a,2} \quad (5)$$

for $i=1,3$, which are the result of photoacoustic tomography experiments with excitation at points 1 and 3. Around point 2 a volume $V_2$ is defined in which a known fraction of photons (here assumed to be unity, although in practice it will be much smaller) that address this volume are 'tagged' or 'labeled'. Assuming an incoming fluence rate $\Phi_{i2}$ and neglecting absorption, the power of labeled photons re-injected in the medium is then

$$P_{L,i2} = \Phi_{i2}A_2 = 4\pi P_i \Pr(i,2)A_2 \quad (6)$$

with $A_2$ the average frontal area of volume 2, with avering over all possible orientations. Here we used Eq. (2) relating the internal fluence rate to the power injected at the medium surface. The internally injected stream of labeled photons at power $P_{L,i2}$ gives rise to detection of labeled photons within an area $A_j$ and solid opening angle $\Omega_j$ at point $j$, at a power that with the use of Eq. (1) can be written

$$P_{L,i2j} = 4\pi P_i \Pr(i,2)\Pr(2,j)A_2 A_j \Omega_j \quad (7)$$

with $(i,j)=(1,3)$ or $(i,j)=(3,1)$.

'Labeling' or 'tagging' of light can be performed by acousto-optic modulation [14, 15] using a focused ultrasound beam, leading to detection of labeled light at the surface of the medium. Here we assume that labeling of light is restricted to volume 2. Following Eq. (7), acousto-optic modulation in volume 2 of light injected in point 1, and detected in point 3, leads to

$$P_{L,123} = 4\pi P_1 \Pr(1,2) \Pr(2,3) A_2 A_3 \Omega_3 \tag{8}$$

By applying $\Pr(2,3) = \Pr(3,2)$, from Eq. (5) with $i=1,3$, and Eq. (8) we obtain three equations that can be solved for the absorption coefficient,

$$\mu_{a,2} = \frac{1}{\Gamma} \sqrt{\frac{A_2 A_3 \Omega_3}{4\pi}} \sqrt{\frac{P_1}{E_{p,1} E_{p,3}}} \sqrt{\frac{\sigma_{21}\sigma_{23}}{P_{L,13}}} \tag{9}$$

This expression for the local absorption coefficient contains only instrumental geometrical parameters (under the first square root), excitation parameters (under the second) and externally measurable quantities (the third square root). The combination of two photoacoustic experiments with injection at two points, and one acousto-optic experiment with the optodes coincident with the photoacoustic injection points leads to the elimination of the probabilities $\Pr(i, j)$ associated with the absorption and scattering properties of the medium. Hence the unknown potentially inhomogeneous optical properties of the tissue, are removed from the problem.

**3. Monte Carlo modeling: method.**

We give numerical evidence of the potential of our method to measure absorption coefficients without knowledge of the local fluence rate or the optical properties of the medium. We used a Monte Carlo simulation based on the program by Jacques and Wang [16, 17], with modifications allowing for tagging photons addressing a certain volume.

The local thermo-elastic stress was replaced by the number of photons absorbed in a predefined sphere around point 2. In this way, simulation of a full photoacoustic experiment is not required. Since in our theory we assumed a tagging efficiency of one, all photons addressing and escaping this sphere were counted as 'tagged'. The sphere is placed in a cubical medium with sides 20mm, having homogenous absorption and scattering properties except for the sphere. The medium is successively illuminated at two surface points for recording the absorption in the sphere, while for tagging only one of these injection points is used, with the other point hosting the light detection window. $10^8$ photons were injected through a circular window of 2mm diameter. The tagged photons were detected in a circular window of 2mm diameter and full opening angle of 50 degrees. In the Monte Carlo simulation the estimation of the absorption coefficient of the sphere is performed with an equivalent of Eq. (9) that reads

$$\mu_{a,2} = \sqrt{\frac{A_3 A_2 \Omega_3}{4\pi V_2^2}} \sqrt{\frac{E_{a,12}^* E_{a,32}^*}{P_{L,3}^*}} \tag{10}$$

with $V_2$ the volume of the absorbing and tagging sphere around point 2, $E_{a,i2}$ is now the number of photons absorbed in the volume, and * denotes normalization with the number of injected photons.

Initially the sphere is placed in the center of the medium, and both the absorption coefficient of the sphere and the bulk are varied. To study the effect of a large variation of fluences, both the depth and the lateral position of the sphere was varied, while the optical properties of the bulk and the sphere were kept constant. Finally, the influence of the size of the sphere on the accuracy of absorption estimation was studied, for a range of reduced scattering coefficients of the sphere and the bulk.

## 4. Results

Fig. 3 shows the estimated vs. the real absorption coefficient, for a range of bulk absorption coefficients within a sphere placed in the center of the medium. The estimation is correct within 8%, with always a positive bias. In this simulation, the local fluence due to variations of the bulk properties varied with a factor 2.2

A more critical test in terms of local fluence variations is to vary the position of the absorbing and tagging volume. In Fig. 4b the estimated absorption coefficient is shown when the volume is shifted from the illumination point at the surface towards the center (black spheres Fig 4a), and from the center towards one of the lateral boundaries (red spheres Fig 4a). The absorption is correctly estimated within 5-10%, with mainly an overestimation except when the absorber is close to the illumination point. Fig. 4c shows the total absorbed energy, equivalent to $\mu_a \Phi$, which varies with a factor of 100. Hence in a normal photoacoustic experiment, the absorber position variation would lead to fluence variations and therefore image value variations of 2 orders of magnitude.

To test the effect of labeling volume size, the absorption has been estimated for labeling volumes of two different sizes placed in the center of the medium, for a range of scattering coefficients of the sphere and the bulk. Fig. 5 shows that the labeling volume has a significant influence on the quality of absorption estimation, with the accuracy increasing with a smaller size of the labeling volume. An underestimation is found for $\mu_s' < 2$ cm$^{-1}$.

## 5. Discussion and conclusion

The method presented here is a double hybrid of photoacoustic imaging and acousto-optic labeling, where the acousto-optic optodes coincide with the two points of injection of pulsed light for photoacoustics. A relationship is derived (eq. 9) that expresses the local absorption coefficient in terms of reconstructed PA image values for two light injection points, the detected power of light labeled in the internal point of interest, and system parameters. Prior knowledge regarding the optical properties of the medium is not required. Simulations give absorption estimations which are correct within a maximum overestimation of 10%, for fluence variations of two orders of magnitude. This dominantly positive bias may have various causes which are related to the size of the labeling volume. Our assumption of an absorbed energy in probing volume V$_2$ equal to $\mu_a \Phi V_2$ only holds if the linear dimensions of the volume are smaller than $1/\mu_a$ and a length scale between $1/\mu_s$ and $1/\mu_s'$. Another source of overestimation is the absorption of tagged photons in the labeling volume. In fact, the fluence suggested by the tagged photons in point 2 (Fig. 2) is lower than the amount of light entering the volume. This missing amount of light is not taken into account in eq. 6.

The simulations suggest that the size of the labeling volume is critical with respect to the effect of the scattering coefficient on the accuracy at which absorption is predicted. The increase of inaccuracy with scattering as shown in Fig. 5 is due to the fact that scattering within the volume significantly increases the path length of photons within the volume, which increases the probability of absorption and hence will lead to an overestimation of absorption. This can be corrected by decreasing the tagging volume as shown in the same figure. Bias of the absorption estimation is also caused by the rather simple tagging model used here, with a unit tagging efficiency for all photons independent of their path length through the labeling volume. A more realistic simulation of the tagging process will lead to tagging proportional to this path length. The underestimation of absorption for low scattering in Fig. 5 is probably caused by a too non-isotropic irradiance for these low scattering levels. The main assumption underlying our theory is that photons injected at 1 and reaching an internal tissue location in a certain direction are proportionally represented by photons injected at 3 and reaching the same location in the opposite direction. This is realized if the irradiances $I_{12}(s)$ and $I_{32}(s)$ in point 2 are each other's point mirrored version, which will generally be the case for multiply scattered light which creates almost isotropic radiances. Close to sources, tissue boundaries or strong inhomogeneities and by shadowing of a neighboring absorbing volume, this condition might break down.

In this study we have proven that the presented method is promising for absolute optical absorption imaging. However, we realize that further investigations are needed to assess the experimental feasibility of the presented technique particularly in vivo. Further work will focus on bridging the gap between the simple discrete labeling sphere with unit tagging efficiency presented in the theory and the simulations, and the physical ultrasound focus which is soft, of complicated shape and with a small tagging efficiency. Both, volume and efficiency, can be affected by many parameters as acoustic attenuation, tissue type at the focal region, coupling of US and tissue and the US modulation in the prefocal and postfocal region. Besides, acousto-optic signals generally suffer from high noise levels. This is especially true in vivo because tissue motion could cause optical speckle decorrelations. Many detection techniques are investigated by other research groups to overcome these problems. Gross et al. [18] had proposed a method to detect tagged photons based on heterodyne parallel speckle detection which allow a high sensitive detection limited only by laser noise (shot noise). Other research groups proposed a detection technique based on photorefractive crystal with fast response time [19], or by applying a quantum spectral filter based on spectral hole burning [20] which allow to overcome speckle decorrelation. These promising techniques have to prove their ability in vivo.

Our method also needs to be elaborated further for more realistic expanded laser beams rather than the assumed collimated narrow beams. The method needs to be tested for its ability to estimate absorption close to tissue boundaries and light sources, and optical inhomogeneities. The presented model assumes the possibility to reconstruct initial stress generated in the medium. But in real tissue this reconstruction is not an easy task. In fact, it depends on many parameters such as acoustic coupling, transducer efficiency, mechanical response, transducer bandwidth and angular aperture, and the area of detection. Therefore, improvement in reconstruction algorithms is needed.

The eventual goal of truly quantitative absorption mapping will allow for quantitative chromophore mapping. This will enable local quantification of natural chromophores such as hemoglobin, e.g. as a result of angiogenic processes or angiogenesis inhibition. Another application is the quantification of targeted contrast agents and the concentration of locally delivered drugs. These applications will be useful in fundamental research, drug development and clinical treatment monitoring.

**Acknowledgements**

This research was supported by the Technology Foundation in the Netherlands (STW) under vici-grant 10831, by Agentschap NL under Eureka grant E!4993, and by the MIRA Institute of the University of Twente. Robert Molenaar and Jithin Jose are acknowledged for experimental support.

**Figures**:

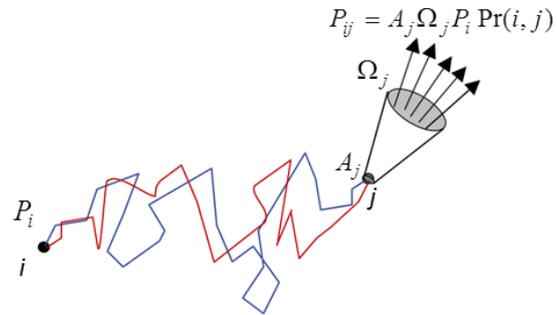

Fig 1. Power $P_{ij}$ measured through aperture with area $A_j$ and solid angle $\Omega_j$ at point $j$ in response to power $P_i$ injected at point $i$. arrows).

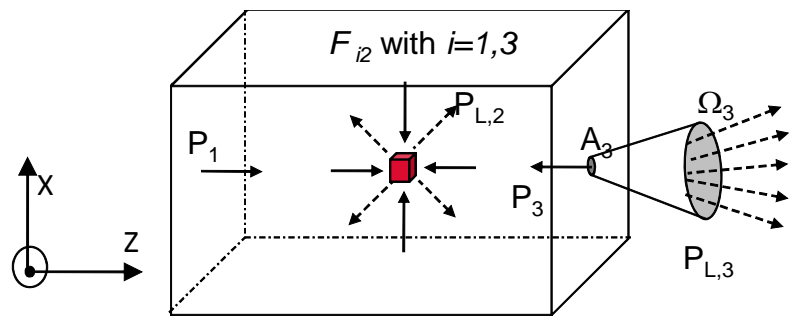

Fig 2. Fluence $\Phi_{i2}$ in internal point 2 in response to light injection in surface points i=1,3 (solid arrows) and detected power of labeled light $P_{L,3}$ in response to acousto-optic modulation at point 2 (dashed arrows).

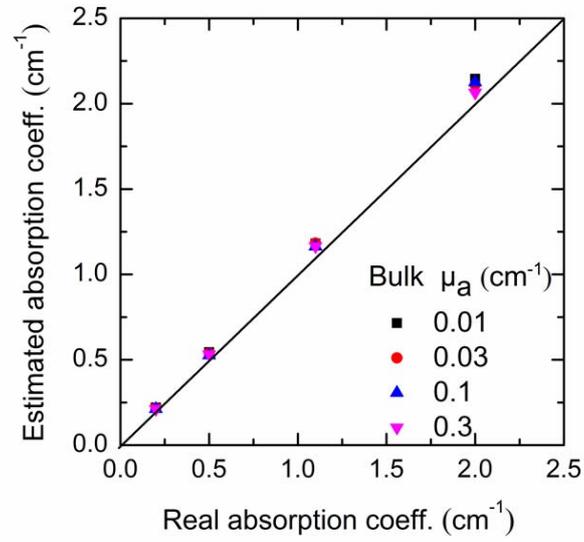

Fig. 3. Estimation of the absorption coefficient in a symmetrically placed absorbing sphere with diameter 2mm, vs. the real absorption coefficient, for a range of bulk absorption levels and a reduced bulk scattering coefficient of 5 cm$^{-1}$, the medium is a 2*2*2 cm$^3$ optically homogeneous object. The solid line represents perfect estimation.

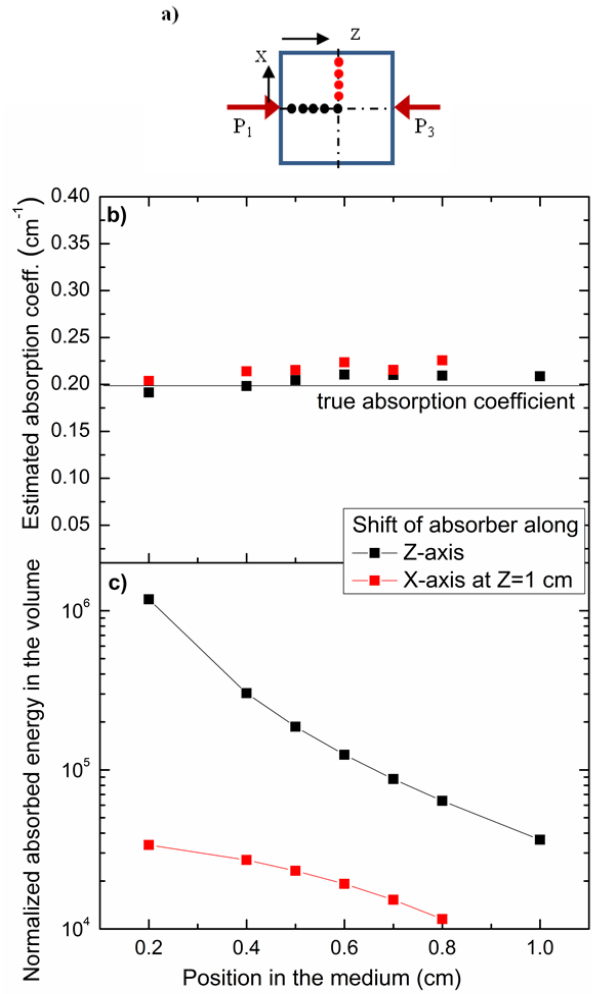

Fig. 4. (a) Schematic of different absorber positions deep in the medium for both directions. (b) Estimated absorption coefficient in 2mm sphere shifting along the x- and z-axis. (c) The associated absorbed energy for injection in point 1, normalized with the number of injected photons, and equivalent to $\mu_a \Phi_a V$ and representing photoacoustic image levels, varies over 2 orders of magnitude.

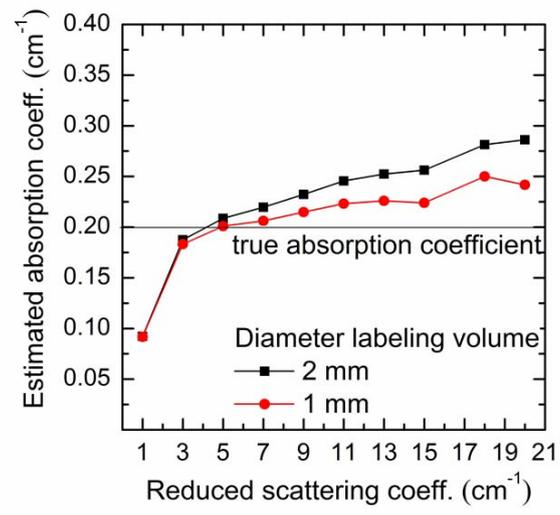

Fig. 5. Estimated absorption coefficient vs. reduced scattering coefficient of bulk and absorbing volume, for labeling spheres with diameters of 2 and 1 mm.